\documentclass[letterpaper,english,reprint,aps,floatfix,prb]{revtex4-2}
\usepackage{amsmath}
\usepackage{amssymb}
\usepackage{graphicx}
\usepackage{array}
\usepackage{dcolumn}
\usepackage{subfigure}
\usepackage{color}
\usepackage{float}
\usepackage{natbib}
\usepackage{epstopdf}
\usepackage[T1]{fontenc}
\usepackage[utf8]{inputenc}
\usepackage[hidelinks=true]{hyperref}
\usepackage[normalem]{ulem}
\usepackage{soul}
\usepackage{bm,stmaryrd}
\usepackage{epstopdf}
\usepackage{multirow}
\usepackage{xcolor}
\usepackage[english]{babel}
\usepackage{hyperref}
\usepackage{textcomp}

\hypersetup{
	colorlinks=true,
	linkcolor=purple,
	citecolor=blue,
	urlcolor=purple
}

\usepackage{babel}

\makeatother

\begin{document}
	\title{Interplay of Valley, Orbital, Spin, and Layer Degrees of Freedom in Ta$_2$CS$_2$ MXene 
    }
\author{Kunal Dutta$^{1}$} \email{pskd2298@iacs.res.in}
\author{Anupam Mondal$^{1}$}
\author{Sayantika Bhowal $^{2}$} \email{sbhowal@iitb.ac.in}
\author{Subhradip Ghosh$^{3}$} \email{subhra@iitg.ac.in}
\author{Indra Dasgupta$^{1}$}  \email{sspid@iacs.res.in} 
\affiliation{$^{1}$ School of Physical Sciences, Indian Association for the Cultivation of Science, 2A and 2B Raja S.C. Mullick Road, Jadavpur, Kolkata 700032, India.\\
	$^{2}$ Department of Physics, Indian Institute of Technology Bombay, Mumbai 400076, India. \\
	$^{3}$ Department of Physics, Indian Institute of Technology Guwahati, Guwahati-781039, Assam, India.}
\date{\today}

\begin{abstract}
We show that the MXene Ta$_2$CS$_2$ provides an excellent platform for hosting multiple coupled degrees of freedom, {\it viz.,} valley, spin, orbital, and layer. The interplay among these degrees of freedom gives rise to a range of intriguing properties in reciprocal space, including valley–orbital and orbital–layer coupling. In the presence of spin–orbit interaction, these couplings lead to valley-dependent and layer-dependent spin splitting of the electronic bands. We further show that the intrinsic electric polarization in Ta$_2$CS$_2$ introduces an additional tuning parameter, enabling control over these coupled degrees of freedom and resulting in switchable valley-dependent orbital moments and Zeeman-like spin splitting. We demonstrate that these nontrivial orbital and spin textures manifest in the orbital and spin Hall effects, respectively. Our results establish noncentrosymmetric MXenes as a promising platform for exploring the interplay among multiple degrees of freedom, their tunability, and the resulting orbital and spin transport phenomena in these two-dimensional materials, thereby paving the way for next-generation spin-orbitronic devices.
\end{abstract}
\maketitle
\section{\protect\label{sec:level1}Introduction}
The orbital Hall effect (OHE) and the spin Hall effect (SHE) are  quantum transport phenomena in which transverse pure orbital and spin currents are generated, respectively, in response to an applied electric field, without any accompanying net charge current~\cite{RevModPhys.87.1213,vonKlitzing2020,PhysRevLett.95.066601,PhysRevLett.123.236403}. These effects have attracted significant attention due to their fundamental importance and their potential applications in next-generation spintronic and orbitronic devices~\cite{RevModPhys.76.323,RevModPhys.80.1517,RevModPhys.91.035004,Cysne2025,Fukami2025,PhysRevLett.99.197202}. Unlike the classical Hall effect, which requires a magnetic field and arises from the Lorentz force acting on charge carriers, the OHE and SHE can occur in nonmagnetic systems without any need for an external magnetic field. In such systems, an external electric field induces a transverse flow of spin or orbital angular momentum, which is typically detected indirectly through spin or orbital injection into an adjacent ferromagnetic layer \cite{PhysRevB.94.020403,PhysRevLett.111.066602,tbdz-jc6k}.
\newline
\indent Unlike the SHE, the OHE  originates from the intrinsic orbital angular momentum of Bloch electrons and does not rely on the spin-orbit coupling (SOC) for its existence. Recent studies have established that nontrivial spin and orbital textures, including Rashba-, Dresselhaus-, and Zeeman-like couplings in both the spin and orbital sectors can generate sizable spin and orbital Hall responses \cite{PhysRevLett.92.126603,PhysRevB.105.245405,doi:10.1073/pnas.2305541120,PhysRevLett.121.086602,PhysRevB.101.075429,PhysRevB.101.161409,pnas_2305541120,PhysRevB.101.121112,PhysRevB.102.035409}.  
A key aspect of the OHE is its ubiquity in multiorbital systems, where multiple orbital degrees of freedom naturally give rise to orbital Berry curvature throughout the Brillouin zone (BZ) \cite{PhysRevLett.121.086602,PhysRevB.103.195309}.
\newline
\indent It is now well established that the OHE is closely linked to the presence of nontrivial orbital textures in momentum space. Prominent orbital Hall responses have been predicted and observed in systems hosting the orbital Rashba effect (ORE)~\cite{PhysRevLett.121.086602}, orbital Dresselhaus effect~\cite{PhysRevB.101.075429,PhysRevB.101.161409}, and radial orbital textures~\cite{pnas_2305541120}, as well as in systems where the wave function at the valleys are eigenfunction of the angular momentum operator~\cite{,PhysRevB.101.121112,PhysRevB.102.035409,PhysRevB.110.054403}. In particular, OHE has been predicted for  group-XIV materials~\cite{PhysRevB.104.245204} and multilayer systems \cite{PhysRevLett.126.056601,PhysRevB.111.165102,PhysRevB.105.195421,PhysRevB.111.L180408}. These orbital textures generate finite orbital Berry curvature, which acts as the driving mechanism for the intrinsic OHE in inversion broken systems. Upon inclusion of SOC, additional transport phenomena, such as the SHE, may emerge, originating from SOC-induced nontrivial spin textures 
~\cite{10.1126_sciadv.aav8575,PhysRevLett.92.126603,PhysRevLett.117.146403}.
\newline
\indent In this context, layered materials are particularly intriguing as they provide an additional degree of freedom, known as the \emph{layer pseudospin}. Electronic states predominantly localized on the upper or lower layer can be identified as pseudospin-up or pseudospin-down states, respectively. In layered systems exhibiting strong spin~, orbital, and valley-dependent couplings, the internal quantum degrees of freedom become entangled with the layer pseudospin. This entanglement gives rise to phenomena such as spin- and orbital-layer locking~\cite{Jones2014}, enabling electrically tunable spin and orbital polarization and providing a versatile platform for exploring multidimensional quantum transport.
\newline
\indent Among the layered materials, transition-metal carbides, nitrides, and carbonitrides, collectively known as MXenes, are a notable family. They have the general chemical formula $M_{n+1}X_nT_x$, where $M$ denotes an early transition metal, $X$ represents carbon and/or nitrogen, $T_x$ refers to surface termination groups, and $n = 1$-$4$ \cite{Li2022}. These materials can be realized as two-dimensional (2D) systems derived from layered $M_{n+1}X_n$ structures through selective etching and subsequent surface functionalization. Depending on the relative positioning of the surface functional groups, MXenes can adopt different structural configurations, commonly classified as X-top, M-top, and mixed configurations \cite{D0NR06609E}. Importantly, a wide range of experimentally accessible functional groups, such as O, OH, F, Cl, and S~\cite{Hart2019,WALTER2000259,SAKAMAKI2002283}, can be employed to functionalize MXene surfaces, resulting in diverse structural phases with rich and tunable electronic properties. 
Interestingly, in the mixed configuration, the functional groups are asymmetrically positioned above the $X$ atoms on one side and above the $M$ atoms on the opposite side of the MXene layer. This asymmetric functionalization breaks inversion symmetry and gives rise to a spontaneous out-of-plane polarization \cite{D0NR06609E}, thereby providing yet another degree of freedom. 
Motivated by this in the present study, we focus on the polar MXene Ta$_2$CS$_2$, where the coexistence of distinct coordination environments and broken inversion symmetry provides an ideal platform for exploring novel electronic phenomena.
\newline
\indent We demonstrate that monolayer Ta$_2$CS$_2$ MXene, realized in two distinct orientations of the electronic polarization, exhibits a strong intrinsic orbital Hall response. Using tight-binding (TB) model calculations along with density functional theory (DFT)-based calculations, we show that the ORE, together with valley-coupled orbital textures, plays a central role in generating a significant orbital Berry curvature, leading to a sizable orbital Hall conductivity (OHC). Upon inclusion of SOC, a non-zero but comparatively weaker spin Hall conductivity (SHC) emerges. The breaking of inversion symmetry and the absence of in-plane mirror symmetry, characterized by the $C_{3v}$ point-group symmetry, confirm the presence of the orbital Rashba effect in both polar phases. Upon introducing an additional layer, a new internal degree of freedom, the layer pseudospin, emerges and couples strongly with both spin and orbital degrees of freedom. This layer-dependent coupling not only enhances the spin Hall effect but also has the potential to significantly modify the orbital Hall response, providing a platform to tune these responses.
\newline
\indent The remainder of the paper is organized as follows. In Sec.~\ref{sec:level2}, we describe the structural details of the system and the computational methodology. Sec.~\ref{sec:level3} presents the main results, including the construction of the tight-binding model, analysis of the orbital Rashba effect, valley-orbital coupling,  calculations of the orbital and spin Hall conductivities, and a detailed study of layer-orbital and layer-spin coupling. Finally, Sec.~\ref{sec:level4} summarizes our findings and presents the conclusions.

\section{\protect\label{sec:level2}Structural and Computational Details}
\subsection{Structural Details}
\begin{figure*}[!t]
\centering
\includegraphics[width= 158 mm ,height=102.0 mm,keepaspectratio]{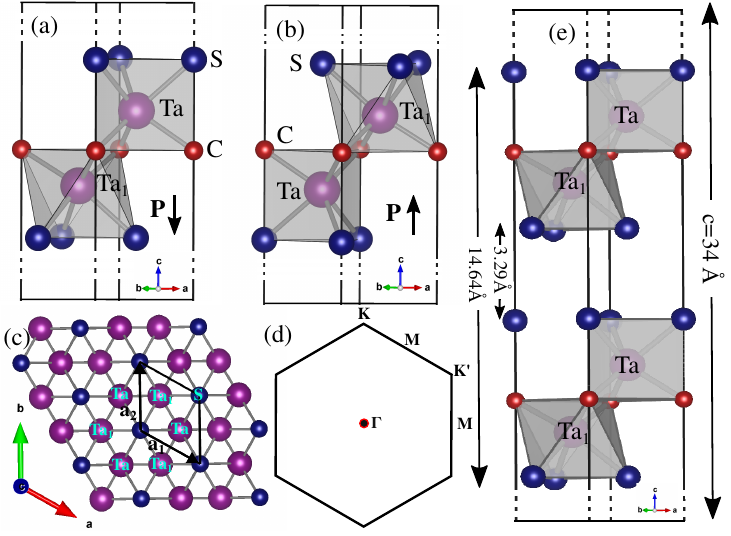}
\caption{Crystal structure of the MXene, Ta$_2$CS$_2$  showing two distinct orientations of the electronic polarization in the monolayer, and the AA stacking configuration in the bilayer. The red, blue, and violet spheres represent C, S, and Ta atoms, respectively. Within the unit cell, the two Ta atoms experience distinct local environments. Panels (a) and (b) depict the monolayer with downward and upward polarization, respectively. Panel (c) shows the top view of the monolayer structure with downward polarization. Panel (d) illustrates the Brillouin zone corresponding to the MXene monolayer with a triangular unit cell. Panel (e) presents the bilayer structure in the AA stacking configuration, where two monolayers with downward polarization are stacked on top of each other.}
\protect\label{fig:figure1}
\end{figure*}
Ta$_2$CS$_2$ MXene can be obtained experimentally by exfoliating its commercially available parent MAX compound, Ta$_2$AlC. During the exfoliation process, surface passivation of MXenes by various functional groups is essentially unavoidable. In particular, removal of the Al layers leads to the formation of S-terminated Ta$_2$CS$_2$ MXene. Experimentally, Ta$_2$CS$_2$ has been reported to crystallize in a trigonal structure with $P3m1$ (No.~156) space group symmetry~\cite{Beckmann1970}. This material is noncentrosymmetric with broken in-plane mirror symmetry and belongs to the $C_{3v}$ point group. Ta$_2$CS$_2$ is intrinsically polar, and two distinct   orientation of electronic polarization can be realized depending on the surface termination generated during exfoliation. In one  electronic polarization, the Ta atoms in the top layer are coordinated by six C/S ligands in a trigonal biprismatic crystal field, while the Ta$_1$ atoms in the bottom layer reside in an octahedral environment, as shown in Fig. \ref{fig:figure1}(a). In the opposite  electronic polarization, the coordination environments are reversed: the Ta atoms in the bottom layer adopt trigonal biprismatic coordination, whereas the Ta$_1$ atoms in the top layer are octahedrally coordinated, as illustrated in Fig~\ref{fig:figure1}(b). 

To investigate multilayer effects, we further construct a bilayer by stacking two monolayers with downward polarization in an AA stacking configuration, such that one downward-polarized layer lies directly on top of the other, as shown in Fig.~\ref{fig:figure1}(e). We confirm the dynamical stability of the structure through explicit phonon calculations (see Sec .~S1 of the Supplementary Information (SI) \cite{supplemental} for details). The resulting bilayer preserves the same crystallographic symmetries as the monolayer, retaining the trigonal space group $P3m1$ (No.~156) symmetry. Consequently, it remains noncentrosymmetric and belongs to the $C_{3v}$ point group. 

The primitive lattice vectors of the system are given by
$\mathbf{a}_1 = \tfrac{\sqrt{3}}{2}a\,\hat{x} - \tfrac{1}{2}a\,\hat{y}$,
$\mathbf{a}_2 = a\,\hat{y}$, 
    where $a= 3.25$~\AA~ is the in-plane lattice parameter. The corresponding BZ of the Ta$_2$CS$_2$ triangular lattice is hexagonal in shape. The high-symmetry points of the BZ, expressed in reciprocal lattice coordinates, include $\Gamma(0,0)$ at the zone center, $K(0,4\pi/3a)$ and $K'(0,-4\pi/3a)$ at the corners of the hexagonal BZ, and $M(\pi/\sqrt{3}a,\,\pi/a)$ at the midpoint of the edge connecting adjacent corners, as shown in Fig. \ref{fig:figure1}(d). Among these points, $\Gamma$ and $M$ are time-reversal-invariant momenta (TRIM), while $K$ and $K'$ are non-TRIM points.
\subsection{Computational Details}
For monolayer Ta$_2$CS$_2$, the DFT~\cite{PhysRev.140.A1133} calculations presented in this work were performed using \texttt{Quantum ESPRESSO}~\cite{Giannozzi_2009}, the Vienna \emph{ab initio} Simulation Package (VASP)~\cite{PhysRevB.54.11169}, and the muffin-tin-orbital-based Nth-order muffin-tin orbital (NMTO) method~\cite{PhysRevB.62.R16219}. Structural optimizations were carried out using VASP until the total energy and the Hellmann–Feynman forces converged to $10^{-6}$~eV and $0.01$~eV/\AA, respectively.

To construct an effective low-energy model for the monolayer retaining only Ta-d states, the TB hopping matrix elements between the Ta-$d$ and Ta$_1$-$d$ orbitals were extracted using the NMTO downfolding method~\cite{PhysRevB.85.014435} keeping up to sixth–nearest-neighbor interactions in real space. The resulting TB Hamiltonian was used to compute the  $\mathbf{k}$-space orbital magnetic moment and the orbital and spin Hall conductivities. The BZ integrations were performed over a $200 \times 200$ $\mathbf{k}$-point mesh in the two-dimensional BZ.

The calculations of the orbital magnetic moment and the SHC for the monolayer were further complemented and for the bilayer were exclusively carried out using  \texttt{Quantum ESPRESSO} and the \texttt{Wannier90} code~\cite{Giannozzi_2009,MOSTOFI20142309}.
The electron–ion interactions in \texttt{Quantum ESPRESSO} were described using optimized norm-conserving Vanderbilt (ONCV) pseudopotentials \cite{PhysRevB.88.085117}, while the exchange–correlation functional was treated within the generalized gradient approximation (GGA) using the Perdew–Burke–Ernzerhof (PBE) functional ~\cite{PhysRevLett.77.3865}. The \emph{ab initio}  ground-state wave functions obtained from self-consistent calculations were employed to construct maximally localized Wannier functions (MLWFs)~\cite{PhysRevB.56.12847}. The wannierization procedure was converged to a tolerance of $10^{-10}$~\AA$^2$, yielding an average Wannier-function spread of approximately $3$~\AA$^2$.

In addition, the VASP code was employed for the calculation of the electronic structure of the bilayer and analysis of orbital-projected band structures. The electron-ion interactions were described using the projector augmented-wave (PAW) method~\cite{PhysRevB.59.1758}, while the exchange-correlation functional was treated within the generalized gradient approximation (GGA) using the Perdew-Burke-Ernzerhof (PBE) functional~\cite{PhysRevLett.77.3865}. A plane-wave kinetic-energy cutoff of $550$~eV was used. To accurately capture inter-layer interactions, van der Waals forces were included for both \texttt{Quantum ESPRESSO} and VASP calculations through the DFT-D3 correction scheme~\cite{PhysRevB.110.155401}.

\section{\protect\label{sec:level3}Results and Discussion}
\begin{figure}[!t]
\centering
\includegraphics[width=89 mm,totalheight=85.0 mm, keepaspectratio,height=130 mm]{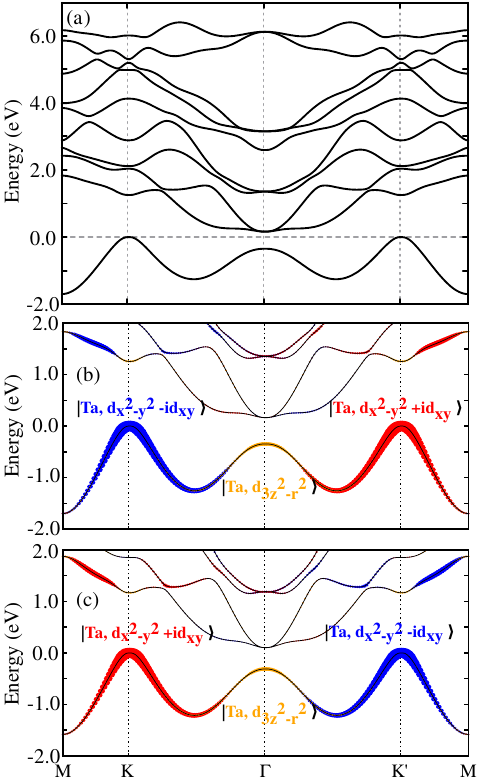}
\caption{ (a) Electronic band structure of a monolayer Ta$_2$CS$_2$ in the nonmagnetic configuration, calculated using the TB model along the high-symmetry path {M}$\left(\frac{\pi}{a\sqrt{3}}, \frac{\pi}{a}, 0 \right)$-K$\left(0, \frac{4\pi}{3a}, 0 \right)$-$\Gamma\left(0, 0\right)$- K$^{\prime}\left( 0, -\frac{4\pi}{3a}, 0 \right)$-M$\left( -\frac{\pi}{a\sqrt{3}}, -\frac{\pi}{a}, 0 \right)$  in the BZ. 
The orbital-projected band structures for (b) downward and (c) upward electric polarization configurations of the monolayer Ta$_2$CS$_2$, highlighting the contributions of Ta-$d$ orbitals obtained from the TB model along the same high-symmetry path. The band structures are shown as black solid lines. The complex orbital combinations $\left(|{\rm Ta}, d_{x^2-y^2}\rangle - i|{\rm Ta}, d_{xy}\rangle\right)$ and $\left(|{\rm Ta}, d_{x^2-y^2}\rangle + i|{\rm Ta}, d_{xy}\rangle\right)$ are represented by blue and red, respectively, while the contribution from the $|{\rm Ta}, d_{3z^2-r^2}\rangle$ orbital is shown in orange. The linear combinations $\left(|d_{x^2-y^2}\rangle \pm i|d_{xy}\rangle\right)$ are eigen-states of the orbital angular momentum operator $L_z$ with eigenvalues $\pm 2\hbar$.} 
    
\protect\label{fig:figure2}
\end{figure}

\subsection{Tight-binding model for the monolayer}

 We begin our discussion with the construction of the TB model for the monolayer Ta$_2$CS$_2$ with downward polarization (see Fig. \ref{fig:figure1}(a)). The total number of valence electrons contributed by Ta (5$d{^3}$6$s^2$), C (5$s{^2}$6$p^2$), and the S (3$s{^2}$3$p^4$) atoms is $26$. For a low-energy description, the completely filled $12$ bands originating from the C and S $s$ and $p$ orbitals are down-folded. Consequently, the effective TB Hamiltonian retains only the Ta-$d$ states, resulting in a total of $10$ bands (excluding spin), which originate from the five $d$ orbitals of each of the two in-equivalent Ta atoms present in the unit cell. The complete orbital basis for the TB Hamiltonian defined on a triangular lattice  consists of ten Ta-$d$ orbitals,
$
	\left| \phi_{n} \right\rangle =
	\bigl(
	|\mathrm{Ta}, xy\rangle, |\mathrm{Ta}, yz\rangle, |\mathrm{Ta}, 3z^2-r^2\rangle,
	|\mathrm{Ta}, xz\rangle, |\mathrm{Ta}, x^2-y^2\rangle,
	|\mathrm{Ta}_1, xy\rangle, |\mathrm{Ta}_1, yz\rangle, |\mathrm{Ta}_1,3z^2-r^2\rangle,
	|\mathrm{Ta}_1, xz\rangle, |\mathrm{Ta}_1, x^2-y^2\rangle \nonumber
	\bigr).
$
To accurately reproduce the low-energy electronic structure obtained from the first-principles calculations, we have included hopping parameters, extending upto the 6-th nearest neighbors. The resulting band structure of the effective TB model  is shown in Fig.~\ref{fig:figure2}(a). As expected, only a single d-band that accommodate remaining two electrons lies below the Fermi level, while the remaining nine bands are located above it, consistent with the Ta-$5d^2$ electronic configuration. The TB band structure closely resembles the corresponding band dispersion obtained from first-principles DFT calculations (see section S2 in the SI \cite{supplemental}). 

We note that the electronic states near the Fermi level in the valence band are dominated by Ta-$d$ orbitals, which experience a trigonal biprismatic crystal-field environment. The corresponding crystal-field splitting~\cite{pph8-2839} determines the orbital character of the bands in the vicinity of the valley points, $K$ and $K^{\prime}$, and $\Gamma$ point in the BZ. The electronic band dispersion of the valence band near the $K$ and $K^{\prime}$ valleys is dominated by the Ta-$d_{x^2-y^2}$ and Ta-$d_{xy}$ orbital characters, whereas the states close to the $\Gamma$ point mainly arise from the Ta-$d_{3z^2-r^2}$ orbital character, consistent with the trigonal prismatic crystal-field environment, under which the Ta $d$ orbitals split into three manifolds: $(d_{x^2-y^2}, d_{xy})$, $(d_{3z^2-r^2})$, and $(d_{xz}, d_{yz})$. Interestingly, at the valley points, the Bloch states form specific linear combinations of the in-plane $d$ orbitals, viz., $\left( |d_{x^2-y^2}\rangle - i |d_{xy}\rangle \right)$ at the $K$ point and $\left( |d_{x^2-y^2}\rangle + i |d_{xy}\rangle \right)$ at the $K^{\prime}$ point, as depicted in Fig.~\ref{fig:figure2}(b). These combinations constitute eigenstates of the orbital angular momentum operator $L_z$, that satisfy $L_z \left( |d_{x^2-y^2}\rangle \pm i |d_{xy}\rangle \right) = \pm 2\hbar \left( |d_{x^2-y^2}\rangle \pm i |d_{xy}\rangle \right)$. 
We find that the electronic band structure of Ta$_2$CS$_2$  with upward polarization (see Fig. \ref{fig:figure1}(b)) is identical to that of the present  structure, except that the orbital angular momentum is reversed at the valley points, as shown in Fig.~\ref{fig:figure2} (c). In contrast, the electronic states in the vicinity of the $\Gamma$ point are predominantly composed of the  $\lvert \mathrm{Ta}, d_{3z^2-r^2} \rangle$.

\subsection{Orbital Rashba Effect (ORE)}\label{ORE}
We next focus on calculating the orbital texture. To this end, we first define the orbital angular-momentum operators in the chosen basis. For the $d$-orbital subspace ($L=2$), we adopt the basis
$\phi_n = \left( d_{xy},\, d_{yz},\, d_{z^2},\,  d_{xz},\, d_{x^2-y^2} \right),$
in which the orbital angular-momentum operators are given by \cite{PhysRevB.110.054403},

\begin{equation}
	L_x^{(d)} = \hbar
	\begin{pmatrix}
			0 & 0 & 0 & -i & 0 \\
			0 & 0 & -i\sqrt{3} & 0 & -i \\
			0 & i\sqrt{3} & 0 & 0 & 0 \\
			i & 0 & 0 & 0 & 0 \\
			0 & i & 0 & 0 & 0 \nonumber
	\end{pmatrix},
\end{equation}

\begin{equation}
	L_y^{(d)} = \hbar
	\begin{pmatrix}
		0 & i & 0 & 0 & 0 \\
		-i & 0 & 0 & 0 & 0 \\
		0 & 0 & 0 & -i\sqrt{3} & 0 \\
		0 & 0 & i\sqrt{3} & 0 & -i \\
		0 & 0 & 0 & i & 0 \nonumber
	\end{pmatrix},
\end{equation}

\begin{equation}
	L_z^{(d)} = \hbar
	\begin{pmatrix}
		0 & 0 & 0 & 0 & 2i \\
		0 & 0 & 0 & i & 0 \\
		0 & 0 & 0 & 0 & 0 \\
		0 & -i & 0 & 0 & 0 \\
		-2i & 0 & 0 & 0 & 0 \nonumber
	\end{pmatrix}.
\end{equation}

For the full TB basis consisting of ten Ta $d$ orbitals (five from each Ta atoms in the unit cell), $\left| \phi_{n} \right\rangle = ( \lvert \mathrm{Ta}, xy \rangle, \lvert \mathrm{Ta}, yz \rangle, \lvert \mathrm{Ta}, 3z^2-r^2 \rangle, \lvert \mathrm{Ta}, xz \rangle, \lvert \mathrm{Ta}, x^2-y^2 \rangle, \lvert \mathrm{Ta}_1, xy \rangle, \lvert \mathrm{Ta}_1, yz \rangle, \lvert \mathrm{Ta}_1, 3z^2-r^2 \rangle, \lvert \mathrm{Ta}_1, xz \rangle, \lvert \mathrm{Ta}_1, x^2-y^2 \rangle ),$ 
the orbital angular-momentum operator has a block-diagonal form,
\begin{equation}
	L_{\alpha} =
	\begin{pmatrix}
		L_{\alpha}^{(d)} & \bf{0} \\
		\bf{0} & L_{\alpha}^{(d)} \nonumber
	\end{pmatrix}.
\end{equation}
Here $\alpha = x, y, z$, and $\bf{0}$ denotes the $5\times5$ null matrix.

Following this, we compute the momentum-resolved orbital angular momentum, defined as,
\begin{equation}
	\mathbf{L}_n(\mathbf{k}) =
	\left\langle u_{n\mathbf{k}} \left| \hat{\mathbf{L}} \right| u_{n\mathbf{k}} \right\rangle.
	\label{eq:equation1}
\end{equation}
Here, the Bloch eigenstates are expressed as
$
	\left| u_{n\mathbf{k}} \right\rangle =
	\sum_{\mu} a^{\mu}_{n\mathbf{k}} \left| \mu \right\rangle,
$ with $\left| \mu \right\rangle$ denoting the atomic $d$ orbitals of the two Ta atoms in the unit cell.

\begin{figure}[!t]
\centering
\includegraphics[width=87mm,totalheight=44.564mm,keepaspectratio,height=7.15cm]{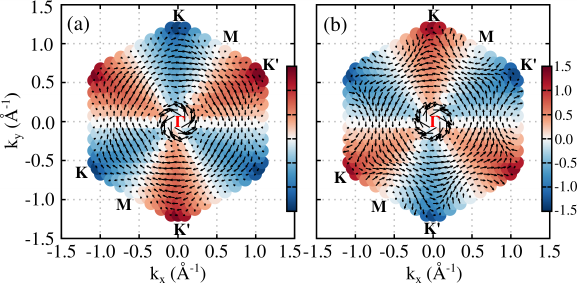}\caption{ Orbital texture of the topmost valence band obtained from the TB model for both structural phases. The in-plane components of the orbital angular momentum expectation values are represented by arrows, while the out-of-plane component is shown using a color scale. Panels (a) and (b) illustrate the orbital textures for the downward and upward polarization states, respectively.
}
\protect\label{fig:figure3}
\end{figure} 

The resulting orbital textures, computed using Eq.~\ref{eq:equation1}, are shown in Fig.~\ref{fig:figure3}(a) for the valence band.
We find that the orbital texture exhibits both in-plane and out-of-plane components.
        The in-plane component displays a chiral winding in reciprocal space, which is known as the ORE \cite{Nikolaev2024}. This behavior is analogous to the conventional spin Rashba effect, where the spin expectation values, plotted around a high-symmetry point, form a chiral winding in momentum space, driven by the SOC. In contrast, in the present case, the orbital texture emerges even in the absence of SOC.

Figures~\ref{fig:figure3}(a) and \ref{fig:figure3}(b) show the orbital textures for the two opposite polarization directions.
Interestingly, we find that the sign of the in-plane components of the orbital moment is reversed upon switching the polarization, demonstrating a direct correspondence between the orbital texture and the direction of electric polarization. We note that our computed orbital texture is consistent with the $C_{3v}$ point-group symmetry of the polar crystal structure. 

Microscopically, the orbital texture originates from strong inter-orbital
hybridization among Ta $d$ orbitals, as evidenced by the inversion symmetry breaking-induced
hopping matrix elements. This hybridization gives rise to finite matrix
elements of the orbital angular momentum operator $\hat{\mathbf{L}}$,
leading to a momentum-dependent orbital texture even in the absence of SOC.
Interestingly, the induced hopping elements further switch their signs for
the opposite polar phase (see section S3 of the SI for more details \cite{supplemental}), leading to the
reversal of the orbital texture between the two polar structures, as seen
from Fig.~\ref{fig:figure3}.


\subsection{Valley-orbital Coupling}
\begin{figure}[!t]
	\centering
	\includegraphics[width=78 mm,totalheight=95 mm,keepaspectratio,height=95 mm]{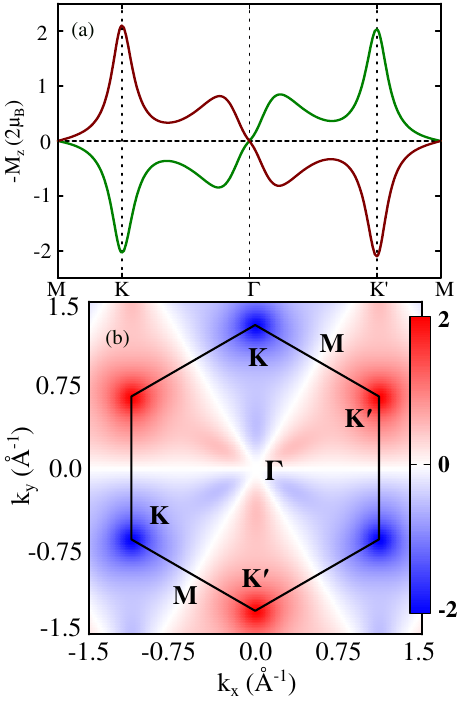}
	\caption{(a) The orbital magnetic moment $M_z(\mathbf{k})$, calculated using the TB model, for the two distinct orientations of the electronic polarization in monolayer Ta$_2$CS$_2$ along a high-symmetry path 
		in the Brillouin zone. The green curve corresponds to the downward electronic polarization, while the maroon curve represents the upward electronic polarization. (b) The distribution of the intrinsic orbital magnetic moment (in units of 2$\mu{_B}$) on the $k_x$-$k_y$ plane of the BZ.}
	\protect\label{fig:figure4}
\end{figure}

We now focus on the out-of-plane component of the orbital angular momentum. As seen from Fig. \ref{fig:figure3}, the out-of-plane component attains its maximum magnitude at the $K$ and $K'$ valleys of the BZ, while their directions are opposite, indicating a pronounced orbital–valley coupling.

To gain further insight, we analyze the out-of-plane component of the orbital moments in the $k$ space using the "modern theory" of the orbital magnetic moment. Within the modern (Berry-phase) theory, the orbital magnetic moment of Bloch electrons is a geometric property of the electronic wave functions, arising from the self-rotation of a Bloch wave packet in real space. Consequently, the orbital magnetic moment $\vec{M}(\vec{k})$ can be computed as \cite{PhysRevLett.95.137204,PhysRevB.74.024408},


\begin{equation}
	\begin{aligned}
		\vec{M}(\vec{k})
		= &  \frac{e}{2\hbar}\,\mathrm{Im}
		\left[
		\langle \nabla_{\vec{k}} u_{\vec{k}} \vert
		\times
		\left( \mathcal{H} - \varepsilon_{\vec{k}} \right)
		\vert \nabla_{\vec{k}} u_{\vec{k}} \rangle
		\right]
		\\
	\end{aligned}
		\label{eq:equation2}
\end{equation}

where $\varepsilon_{\mathbf{k}}$ and $u_{\mathbf{k}}$ are the energy eigenvalues and the cell-periodic parts of the Bloch eigenfunctions for a given band, respectively. The total orbital magnetic moment is obtained by summing over all occupied states at each $\vec{k}$-point in the BZ.

Our computed ${M}_z(\vec{k})$ for the valence band of the TB model, obtained using Eq.~\ref{eq:equation2}, is shown in Fig. \ref{fig:figure4}(a) along the high-symmetry $k$-path in the BZ. As seen from Fig. \ref{fig:figure4}(a), ${M}_z(\vec{k})$ has the same magnitude but opposite sign at the valley points, consistent with our atom-centered results, shown in Fig. \ref{fig:figure3}. This behavior can be understood from the transformation of the orbital moment under the inversion ($\mathcal{P}$) and time-reversal ($\mathcal{T}$) symmetries: $\mathcal{P}\, \vec  M(k_{x}, k_{y}) = \vec M(-k_{x}, -k_{y})$
and
$\mathcal{T}\, \vec M(k_{x}, k_{y}) = - \vec M(-k_{x}, -k_{y})$. Consequently, the simultaneous presence of $\mathcal{P}$ and $\mathcal{T}$ symmetries enforces $\vec M(k_{x}, k_{y}) = 0$ for all $\vec{k}$. In the present case, the $\mathcal{P}$ symmetry is broken while the $\mathcal{T}$ symmetry is preserved, leading to a non-zero orbital moment in $k$ space. Furthermore, as follows from the transformation relations, the presence of $\mathcal{T}$ symmetry enforces the orbital moment at the two valleys to be opposite in sign, explaining the computed results in Fig. \ref{fig:figure4}(a).

\begin{figure}[t]
	\includegraphics[height= 100 mm, width=72 mm,keepaspectratio]{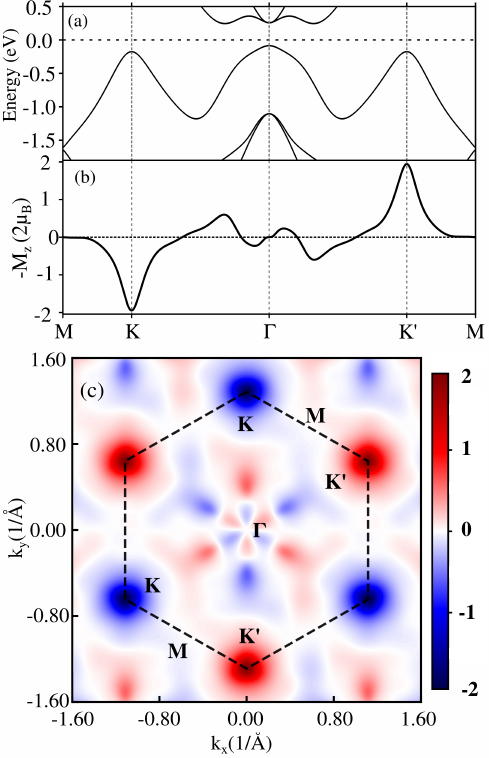}
	\caption{(a) The electronic band structure of monolayer Ta$_2$CS$_2$ with downward electric polarization, calculated with \texttt{Quantum ESPRESSO} and \texttt{Wannier90}, shown along a high-symmetry $k$-path 
in the BZ. (b) The orbital magnetic moment $M_z(\mathbf{k})$, plotted along the same high-symmetry path.
(c) The distribution of the intrinsic orbital magnetic moment (in units of 2$\mu{_B}$) on the $k_x$-$k_y$ plane of the BZ. }
	\label{fig:figure5}
\end{figure}
Furthermore, our calculations for the other structural orientation show that the orbital magnetic moment at each momentum point is reversed, i.e., it has nearly the same magnitude but the opposite sign throughout the BZ, as shown in Fig. \ref{fig:figure4}(a). This suggests that, similar to the ORE, valley-orbital coupling is also switchable upon reversal of the polar distortion. 

Using Eq.~\ref{eq:equation2}, we compute the orbital moment in the $k_x$-$k_y$ plane of the Brillouin zone (BZ) for the structure with downward polarization. The resulting distribution of the orbital magnetic moment, shown in Fig.~\ref{fig:figure4}(b), exhibits pronounced peaks near the valley points $K$ and $K^\prime$, reaching values as large as $\sim 4\,\mu_B$. 

We further complement our model calculation of the orbital magnetic moment \cite{orbital_moment_qe} in Ta$_2$CS$_2$ using \textsc{Quantum ESPRESSO} in combination with the Wannier90 interpolation and post-Wannier90 tools . Here, we derive an effective model that includes the C ($p$) and S ($p$) orbitals in addition to the Ta ($d$) states. Consequently, the effective TB Hamiltonian includes a total of $19$ bands (excluding spin), arising from the five $d$ orbitals of each of the two inequivalent Ta atoms (Ta and Ta$_1$) in the unit cell, together with three C ($p$) orbitals and six S ($p$) orbitals.  The resulting band structure and the $\mathbf{k}$-space orbital moment \cite{PhysRevB.74.024408,PhysRevB.85.014435}, evaluated along high-symmetry lines and over the $k_x$-$k_y$ plane, are shown in Fig.~\ref{fig:figure5}(a)-(c) respectively.   We note that these results are in good agreement with those obtained from our Ta-$d$-only TB model calculations, further confirming the existence of the $k$-space orbital texture in monolayer Ta$_2$CS$_2$.  The differences in the pattern around the $\Gamma$ point of the orbital moment may be attributed to differences in the basis sets of the two TB models and the expression for the orbital magnetic moment. Fig. \ref{fig:figure5}(c) displays the distribution of orbital magnetic moment $M_z$ in the entire BZ confirms equal and opposite values at $K$ and $K^{\prime}$ points.

\subsection{Orbital Hall Effect (OHE)}
\subsubsection{Atom Center Approximation (ACA)}
Since nontrivial orbital textures can give rise to the OHE \cite{PhysRevLett.121.086602}, we now focus on investigating  OHE in  Ta$_2$CS$_2$.
The transport of orbital angular momentum, manifested through phenomenon such as the OHE, forms the central theme of orbitronics. The orbital degree of freedom plays a role analogous to spin in conventional spintronics. Consequently, understanding the orbital transport properties is essential for the development of orbital-based electronic devices, i.e., orbitronic devices.

The orbital magnetic moment and the OHE share a common geometric origin, both arising from the momentum-space structure of Bloch wave functions. In particular, as we found earlier, the state-resolved orbital magnetic moment is strongly enhanced near the inequivalent valley points $K$ and $K^{\prime}$, where the Bloch states carry a well-defined orbital angular momentum and behave approximately as eigenstates of the $L_z$ operator. Motivated by this pronounced valley–orbital character, we investigate the intrinsic orbital Hall response by explicitly evaluating the orbital Berry curvature in momentum space, which is central to the OHE.

We compute the orbital Berry curvature of monolayer Ta$_2$CS$_2$ within the Kubo formula (linear response theory). For the $n$th band, the orbital Berry curvature is defined as
\begin{equation}
	\Omega^{\gamma,\mathrm{orb}}_{n,\alpha\beta}(\mathbf{k}) = 
	2\hbar \sum_{n' \ne n} 
	\mathrm{Im} \left[
	\frac{
		\langle u_{n\mathbf{k}} | J^{\gamma,\mathrm{orb}}_{\alpha} | u_{n'\mathbf{k}} \rangle
		\langle u_{n'\mathbf{k}} | v_{\beta} | u_{n\mathbf{k}} \rangle
	}{
		\left( \varepsilon_{n\mathbf{k}} - \varepsilon_{n'\mathbf{k}} \right)^2
	}
	\right],
\label{eq:equation3}
\end{equation}
where $\varepsilon_{n\mathbf{k}}$ and $u_{n\mathbf{k}}$ denote the eigenvalues and eigenstates of the Hamiltonian, respectively. The orbital current operator is given by
$J^{\gamma,\mathrm{orb}}_{\alpha} = \frac{1}{2}\{v_{\alpha}, L_{\gamma}\}$,
with $L_{\gamma}$ being the orbital angular momentum operator, defined in section \ref{ORE} (see Eq. \ref{eq:equation1}), and
$v_{\alpha} = \frac{1}{\hbar} \partial H / \partial k_{\alpha}$ is the velocity operator.
\begin{figure}[!t]
	\centering
	\includegraphics[width=85mm,totalheight=47mm,keepaspectratio,height=4.7cm]{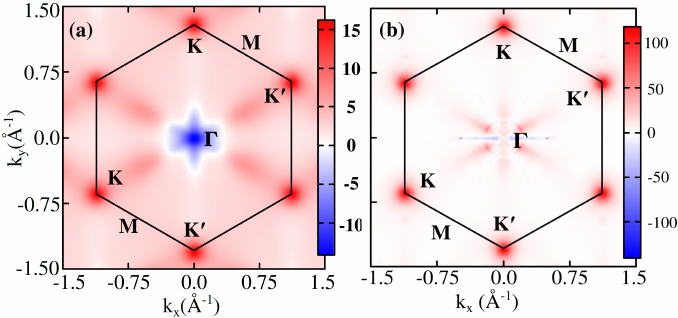}\caption{  (a) orbital Berry curvature  (in units of \AA$^{2}$) on the $k_x$-$k_y$ plane for the lowest occupied band of the TB Hamiltonian, computed using Eq.~\ref{eq:equation3}, where the orbital magnetic moment is obtained from Eq.~\ref{eq:equation1}. (b)  Total orbital Berry curvature  (in units of \AA$^{2}$) on the $k_x$-$k_y$ plane for the lowest occupied band of the TB Hamiltonian, computed using Eq.~\ref{eq:equation3}, where the orbital magnetic moment is obtained from Eq.~\ref{eq:equation5}. The hexagon indicates  the first Brillouin zone.}
	\protect\label{fig:figure6}
\end{figure}
Fig.~\ref{fig:figure6}(a) shows the momentum-space distributions of the orbital Berry curvature, obtained from Eq.~\ref{eq:equation3}, for the lowest occupied band on the $k_x$–$k_y$ plane, using the TB model. We note that, similar to the orbital moment, orbital Berry curvature exhibits large values around the valley points. While the orbital magnetic moment changes sign between $K$ and $K^{\prime}$, reflecting valley-contrasting orbital polarization, the orbital Berry curvature retains the same sign at both valleys. This distinct behavior leads to a nonvanishing net orbital Hall response when the orbital Berry curvature is integrated over the BZ.

Within the linear response theory, an applied electric field $E_k$ induces an orbital current density according to $J^{i}_{j} = \sigma^{i}_{jk} E_{k},$
where $J^i_j$ represents an orbital current flowing along direction $j$ with orbital polarization along the $i$ direction. The corresponding OHC tensor $\sigma^{i}_{jk}$ is obtained by integrating the orbital Berry curvature over all occupied states,
\begin{equation}
	\sigma_{\alpha\beta}^{\gamma,\mathrm{orb}}
	= -\frac{e}{N_k V_c} 
	\sum_{n\mathbf{k}}^{\mathrm{occ}}
	\Omega_{n,\alpha\beta}^{\gamma,\mathrm{orb}}(\mathbf{k}),
	\label{eq:equation4}
\end{equation}
where $N_k$ is the number of $\mathbf{k}$ points and $V_c$ is the unit-cell volume. For the two-dimensional system considered here, $V_c = \frac{\sqrt{3}}{2}a^2$ corresponds to the surface unit-cell area, and the OHC is expressed in units of $(\hbar/e)\,\Omega^{-1}$.

We note that the crystal symmetry imposes strong constraints on the allowed tensor components. For the trigonal space group $P3m1$ (No.~156), symmetry permits only the antisymmetric components, $\sigma_{xy}^{z,\mathrm{orb}} = -\sigma_{yx}^{z,\mathrm{orb}}$, to be nonzero. As seen from Fig. \ref{fig:figure6}(a), along with a large orbital Berry curvature near the valley points, a sizable contribution with opposite sign is also present in the vicinity of the $\Gamma$ point. Nevertheless, over most part of the BZ, the orbital Berry curvature remains predominantly positive. Summing $\Omega_{xy}^{z,\mathrm{orb}}$ over the occupied bands using Eq.~\ref{eq:equation4}, therefore, leads to a large intrinsic OHC, $\sigma^{z,\mathrm{orb}}_{xy} = -1.45 \times 10^{4}\, (\hbar/e)\,\Omega^{-1}$ (without including the spin degeneracy).

\subsubsection{Modern theory}
Within the ACA, only the local contribution of the orbital magnetic moment is retained, while the itinerant (nonlocal) contribution is neglected. 
Therefore, we further evaluate the OHE by incorporating both local and nonlocal contributions to the orbital angular momentum (OAM)~\cite{PhysRevLett.95.137205,PhysRevLett.99.197202}. The symmetrized form of the local OAM operator is given by $\hat{\mathbf{L}} = \frac{1}{4} \left( \hat{\mathbf{r}} \times \hat{\mathbf{p}} - \hat{\mathbf{p}} \times \hat{\mathbf{r}} \right).$ The matrix elements of the OAM operator between Bloch states can be expressed in terms of the velocity operator as \cite{PhysRevB.103.195309,doi:10.1142/S0217979211058912,PhysRevB.106.104414}
\begin{equation}
	\begin{aligned}
		\langle u_{n\mathbf{k}} | \hat{L} | u_{p\mathbf{k}} \rangle 
		= &\frac{e \hbar^2}{4 g_L \mu_B} \, \mathrm{Im} 
		\sum_{q \neq n,p}
		\left(
		\frac{1}{\epsilon_{q\mathbf{k}} - \epsilon_{n\mathbf{k}}}
		+ \frac{1}{\epsilon_{q\mathbf{k}} - \epsilon_{p\mathbf{k}}}
		\right) \\
		& \times 
		\langle u_{n\mathbf{k}} | \hat{\mathbf{v}} | u_{q\mathbf{k}} \rangle 
		\times 
		\langle u_{q\mathbf{k}} | \hat{\mathbf{v}} | u_{p\mathbf{k}} \rangle,
	\end{aligned}
	\label{eq:equation5}
\end{equation}
where $|u_{n\mathbf{k}}\rangle$ denotes the periodic part of the Bloch wavefunction with band energy $\epsilon_{n\mathbf{k}}$, and $\hat{\mathbf{v}} = \hbar^{-1} \partial_{\mathbf{k}} \mathcal{H}_{\mathbf{k}}$ is the velocity operator. Here, $g_L = 1$ for transition metals \cite{AHMacDonald_1982}, and $\mu_B = \frac{e\hbar}{2m_e}$ is the Bohr magneton. It is interesting to note taking the diagonal component ($n = p$) in the above expression, we recover the result of the modern theory for the orbital magnetic moment of a given band, as discussed previously in Eq.\ref{eq:equation2}. However, a complete description of the OHE requires both diagonal ($n = p$) and off-diagonal ($n \neq p$) contributions.  

Based on the constructed OAM operator Eq.~\ref{eq:equation5}, we have evaluated the orbital current operator and subsequently compute the orbital Berry curvature in the $k_x$-$k_y$ plane, as shown in Fig.~\ref{fig:figure6}(b). The resulting orbital Berry curvature, obtained by including both local and nonlocal contributions, closely resembles the ACA results near the valley points K and K$^\prime$, as discussed earlier, with only minor deviations around the $\Gamma$ point. Finally, we compute the OHC as $\sigma^{z,\mathrm{orb}}_{xy} = -2.54 \times 10^{4}\, (\hbar/e)\,\Omega^{-1},$
excluding spin degeneracy. This value is significantly larger than that obtained within the ACA, highlighting the crucial role of nonlocal contributions, suggesting Ta$_2$CS$_2$ as a promising candidate for orbitronic applications.
\begin{figure*}[!t]
\centering
\includegraphics[width=130.0 mm,totalheight=102.0 mm,keepaspectratio,height=102.0 mm]{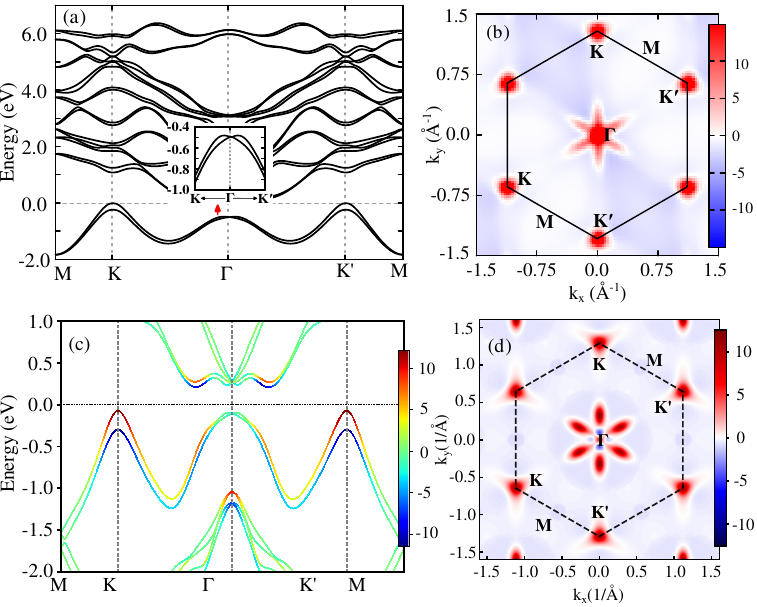}
\caption{(a) TB band structure calculated from the model Hamiltonian with SOC strength $\lambda = 140~\text{meV}$, plotted along a high-symmetry $k$-path in the BZ. The inset shows a magnified view near the $\Gamma$ point, highlighting the Rashba splitting.
(b) Spin Berry curvature distribution (in units of $\text{\AA}^2$) in the $k_z = 0$ plane at the Fermi energy, as indicated by the dashed horizontal line in panel (a).
(c) Band-resolved spin Berry curvature plotted along the same high-symmetry path, calculated using \texttt{Quantum ESPRESSO} and \texttt{Wannier90}.
(d) Corresponding spin Berry curvature distribution (in units of $\text{\AA}^2$) in the $k_z = 0$ plane, summed over all occupied bands.}
\protect\label{fig:figure8}
\end{figure*}
\subsection{Spin-valley coupling and spin Hall effect (SHE)}\label{SHE}
The  OHE, as discussed above, occurs even without including the effect of  SOC primarily because the orbital current originates from the orbital degree of freedom, which does not require SOC for its existence. However, in order to capture the SHE,  it is essential to include SOC explicitly in the Hamiltonian.



After including the SOC term in the TB model, we tune the SOC strength $\lambda$ to reproduce the spin splitting at the valley (K and K$^{\prime}$) points as obtained from the  DFT calculation. The spin splitting at the valleys is approximately $2\lambda$, and with $\lambda = 140$~meV, we find a reasonable agreement between the band structure obtained from the model Hamiltonian and the \emph{ab initio} DFT results for Ta$_2$CS$_2$.  The resulting band structure from the model Hamiltonian is shown in Fig.~\ref{fig:figure8} (a).

As shown in Fig.~\ref{fig:figure8} (a), there is a Zeeman-like splitting of bands caused by SOC at the valley points, K and K$^\prime$, while in the vicinity of the $\Gamma$ point a Rashba-like spin splitting emerges (see the inset of Fig.~\ref{fig:figure8} (a)). This behavior is in good agreement with that reported in Ref. \cite{pph8-2839} and further indicates that both crystallographic environment and the breaking of inversion-symmetry play crucial roles in shaping the spin-orbit driven band structure in this system.

From the energy eigenvalues and eigenfunctions  calculated from  the TB model in the presence of SOC, we compute the spin Berry curvature. The spin Berry curvature is analogous to that of the orbital Berry curvature given in Eq.~\ref{eq:equation3}, except that the orbital current operator $J^{\gamma,\mathrm{orb}}_{\alpha}$ is replaced by the spin current operator $	J^{z,\mathrm{spin}}_{\alpha} = \frac{1}{4}\{v_{\alpha}, s_z\}, $ where $s_z$ is the Pauli matrix corresponding to the $z$ component of the spin.

We compute the spin Berry curvature by adding up  individual contributions of both valence bands, and the result is shown in Fig.~\ref{fig:figure8}(b). Similar to the orbital Berry curvature, the dominant contributions arise from the valley points as well as from the vicinity of the $\Gamma$ point. However, in contrast to the orbital case, the spin Berry curvature is significantly smaller in magnitude. This is understood from the fact that the spin-up and spin-down bands contribute with opposite signs to the spin Berry curvature. However, the magnitudes of these contributions are not exactly identical due to the broken $\mathcal{I}$ symmetry, resulting in a small net intrinsic spin Hall response.

We compute the SHC by summing the spin Berry curvature over the occupied part of the BZ, $	\sigma^{z,\mathrm{spin}}_{xy}
	= -\frac{e}{N_k V_c} 
	\sum_{n\mathbf{k}}^{\mathrm{occ}}
	\Omega^{z,\mathrm{spin}}_{n,xy}(\mathbf{k}). $
The resulting value of the SHC is  $\sigma^{z,\mathrm{spin}}_{xy} =1.01 \times 10^2 \frac{\hbar}{e}\Omega^{-1} $, which is much smaller than the corresponding value of the OHC. Furthermore, we find that the SHC takes nearly the same value in both structural phases, indicating the weak dependence of the SHC on the electric polarization direction. 

We have also calculated the SHE using \textsc{Quantum ESPRESSO} in 
combination with Wannier90 interpolation and post-Wannier90 analysis \cite{PhysRevB.98.214402,PhysRevLett.100.096401}. The computed band-resolved spin Berry curvature along a high-symmetry $k$ path is shown in Fig.~\ref{fig:figure8}(c). The distribution of the spin Berry curvature 
on the $k_x$-$k_y$ plane, summed up to the topmost valence band, is also shown in 
Fig.~\ref{fig:figure8}(d). The corresponding SHC is found to be
$\sigma^{z,\mathrm{spin}}_{xy} = 1.49 \times 10^2 \frac{\hbar}{e} \Omega^{-1}$. We note that these results are in reasonable agreement with those obtained from our Ta-$d$-only TB model Hamiltonian, validating further our model calculations \cite{PhysRevB.106.104414}. 


\begin{figure}[!t]
	\centering
	\includegraphics[width=75 mm,totalheight=79 mm,keepaspectratio,height=7.9 cm]{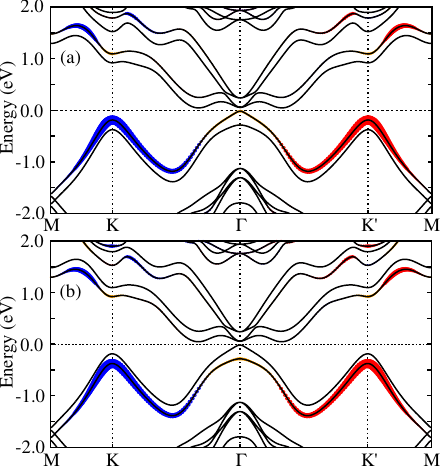}\caption{ Band structure (black lines) in the absence of SOC along with orbital projection along a high-symmetry path.  
    The Fermi level is set to zero on the energy axis. The color-coded orbital characters correspond to $\mathrm{Ta}(d_{x^2-y^2} + i d_{xy})$ (red),  $\mathrm{Ta}(d_{x^2-y^2} - i d_{xy})$ (blue), and  $\mathrm{Ta}(d_{z^2})$ (orange). Panels (a) and (b) show the orbital projections originating from the lower and upper layers, respectively.}
	\protect\label{fig:figure9}
\end{figure}

\subsection{Orbital layer Coupling}
\begin{figure}[!t]
\centering
\includegraphics[height= 112 mm, width=80 mm,keepaspectratio]{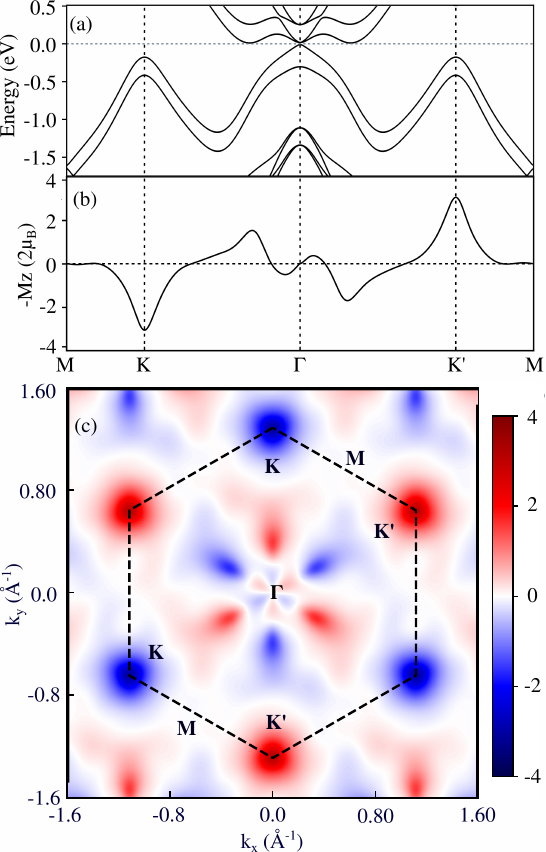}
\caption{(a) The electronic band structure of bilayer Ta$_2$CS$_2$, calculated with \texttt{Quantum ESPRESSO} and \texttt{Wannier90} along a high-symmetry $k$ path. (b) The corresponding orbital magnetic moment $-M_z(\mathbf{k})$, computed along the same high-symmetry path.
(c) The distribution of the intrinsic orbital magnetic moment (in units of $2\mu{_B}$) over the $k_x$-$k_y$ plane of the BZ. }
\label{fig:figure10}
\end{figure}

So far, we have discussed the properties of the monolayer Ta$_2$CS$_2$, where most of the salient features originate from the valley degrees of freedom with well-defined orbital character. We now extend our analysis to the bilayer system.

In the bilayer structure, the two monolayers are stacked in an AA configuration.
The band structure for the bilayer is calculated using VASP and as well as \texttt{Quantum ESPRESSO} and they are in good agreement. By comparing the band structure of the bilayer system, as shown in Fig.~\ref{fig:figure9}, with that of the monolayer, we find noticeable deviations in the conduction band near the $\Gamma$ point, indicating appreciable interlayer hybridization and band mixing in this region. In contrast, the band dispersion around the $K$ and $K^{\prime}$ valleys remains essentially unchanged compared to the monolayer case. This indicates that the electronic structure near the valley points is, to a very good approximation, described by a simple superposition of the two monolayer band structures. 
\begin{figure*}[!t]
\centering
\includegraphics[width=180mm,totalheight=44.564mm,keepaspectratio,height=7.15cm]{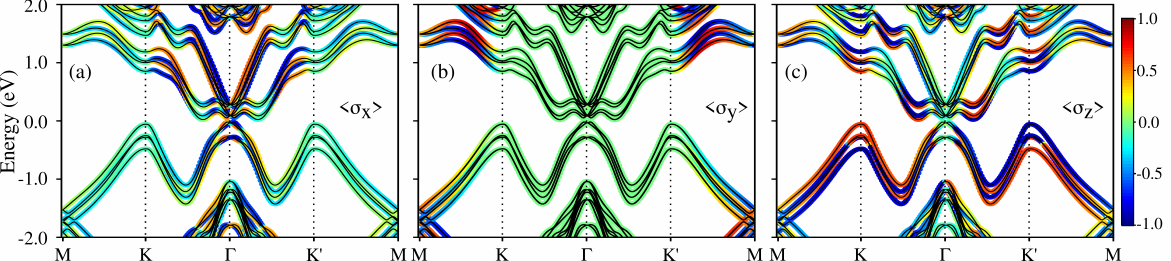}\caption{Expectation values of the spin components,  (a) $\langle \sigma_x \rangle$, (b) $\langle \sigma_y \rangle$, and (c) $\langle \sigma_z \rangle$, projected onto the band structure of the bilayer Ta$_2$CS$_2$ in the presence of SOC along a high-symmetry path. 
The color scale represents the magnitude and sign of the spin expectation values.}
\protect\label{fig:figure11}
\end{figure*}
This is also evident from our analysis of the orbital-projected band structure, as shown in Fig.~\ref{fig:figure9}(a) and (b), where the valley eigenstates $\mathrm{Ta}(d_{x^2-y^2} \mp i d_{xy})$ of the two Ta layers are projected onto the respective layers. We find that the valley states originating from each layer remain well separated in momentum space, while preserving their orbital character. Consequently, the Bloch wavefunctions at the $K$ and  $K^{\prime}$ points remain eigenstates of the orbital angular momentum operator $L_z$ within each layer, leading to a clear orbital-layer locking at the valley points. This further confirms that the interlayer coupling is strongly suppressed in the valley region, while being substantial around the $\Gamma$ point. Such a selective hybridization provides a natural explanation for the robust valley physics in the bilayer system and its strong connection to the orbital degree of freedom. 

An interesting consequence of the orbital-layer coupling is an enhanced orbital 
magnetic moment around the $K$ and $K^\prime$ points, which is expected to be 
nearly twice as large as that of the monolayer. This becomes evident from our DFT calculations of the orbital magnetic moment for the bilayer system. For the DFT calculations, we used the same basis states as described earlier for the monolayer case, leading to a total of 38 bands (excluding spin) due to the doubling of the layer index.
The resulting band structure and the $\mathbf{k}$-space orbital moment, computed along a high-symmetry $k$ path and over the $k_x$-$k_y$ plane, are shown in Fig.~\ref{fig:figure10}(a)-\ref{fig:figure10}(c), respectively. As seen from Fig. ~\ref{fig:figure10}(b) and (c),  the orbital moment in the bilayer system is significantly enhanced and approximately doubled compared to the monolayer case, confirming the presence of orbital-layer coupling in the system.

\subsection{Spin layer Coupling }

After discussing the orbital-layer coupling, we now show that in the presence of SOC, the orbital-layer coupling translates into spin-layer coupling. This is analogous to the spin-valley coupling, which results from orbital-valley coupling in the monolayer and gives rise to spin splitting of bands at the $K$ and $K^{\prime}$ points of the BZ with opposite spin expectation values in the two split branches (see section \ref{SHE}).  

A similar behavior is observed in the bilayer. The spin-resolved band structures, shown in Fig.~\ref{fig:figure11}(a)-(c), indicate that at a given valley, the upper and lower spin-split bands originating from both layers exhibit identical spin polarization following the orbital polarization of the bilayer, as depicted in Fig. \ref{fig:figure9}. We note that similar to the monolayer, the bilayer system lacks inversion symmetry, which plays a crucial role in generating the spin splitting.
Since the $K$ and $K^{\prime}$ valleys are related by time-reversal symmetry, the spin polarizations of the upper and lower spin-split bands of each layer are interchanged between the two valleys, as shown in Fig.~\ref{fig:figure11}(a)-(c). 
 
These results suggest that in reciprocal space, stacking two monolayers with identical orientations maps the $K$ ($K^{\prime}$) points of the individual layers onto the corresponding $K$ ($K^{\prime}$) point of the combined BZ. As a consequence, the valley degrees of freedom of the two layers are directly superimposed, leading to a combined valley response in the bilayer system. The electronic states, therefore, exhibit two distinct types of band splitting: one originating from SOC and the other arising from interlayer coupling. This double splitting is most pronounced around the $K$ and $K^{\prime}$ points (see Fig.~\ref{fig:figure11}(c)), where the low-energy electronic states can be described by the following minimal effective Hamiltonian ~\cite{PhysRevMaterials.8.024005}:
\begin{equation}
\label{eq:equation6}
	H_K = -\frac{\Delta_{\mathrm{spin}}}{2} \, \sigma_z 
	- \frac{\Delta_{\mathrm{layer}}}{2} \, \tau_z ,
\end{equation}
where $\sigma_z$ and $\tau_z$ are Pauli matrices representing the spin and layer pseudospin degrees of freedom, respectively. Here $\Delta_{\mathrm{spin}}$ and $\Delta_{\mathrm{layer}}$ quantify the strengths of the SOC-induced and interlayer-induced splitting. We extract the energy splittings induced by the interlayer coupling and the SOC from the DFT band structure of the bilayer system. The obtained values are approximately 
$97.3$~meV and $113.5$~meV, respectively. This indicates that the spin-orbit-induced splitting is larger than that arising from the interlayer hybridization.

\begin{figure}[!t]
\includegraphics[height= 105 mm, width=87 mm,keepaspectratio]{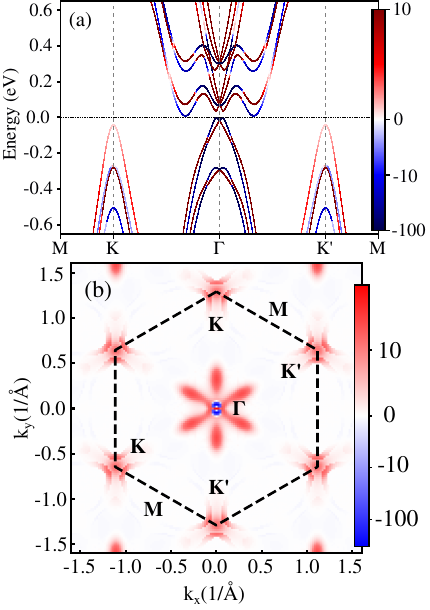}
\caption{ (a) Band-resolved spin Berry curvature along a high-symmetry $k$-path in the BZ, calculated with \texttt{Quantum ESPRESSO} and \texttt{Wannier90} for the bilayer Ta$_2$CS$_2$. (b) The corresponding spin Berry curvature distribution (in $\text{\AA}^2$) in the $k_z = 0$ plane, summed over all occupied bands. The color map shows the value of the spin Berry curvature on a log scale. 
}
\label{fig:figure12}
\end{figure}

In the bilayer configuration, the simultaneous presence of spin-valley and spin-layer coupling at the valley points significantly enhances the spin Hall response, as evident from our results, using \textsc{Quantum ESPRESSO} in combination with Wannier90 
interpolation and post-Wannier90 analysis for bilayer Ta$_2$CS$_2$.
The calculated band-resolved spin Berry curvature along the high-symmetry path 
M-K-$\Gamma$-K$^\prime$-M is shown in Fig.~\ref{fig:figure12}(a). The corresponding
distribution of the spin Berry curvature in the $k_x$-$k_y$ plane, summed up to 
the topmost valence band, is shown in Fig.~\ref{fig:figure12}(b). As evident from these figures, the magnitude of the spin Berry curvature has enhanced significantly in the bilayer system compared to that of the monolayer. These results demonstrate that the interplay between valley, orbital, spin, and layer degrees of freedom provides an effective knob for tuning and enhancing the SHE in bilayer structures.

\section{\protect\label{sec:level4}Summary and Outlook}

\begin{figure}[t]
\includegraphics[height= 75 mm, width=86 mm,keepaspectratio]{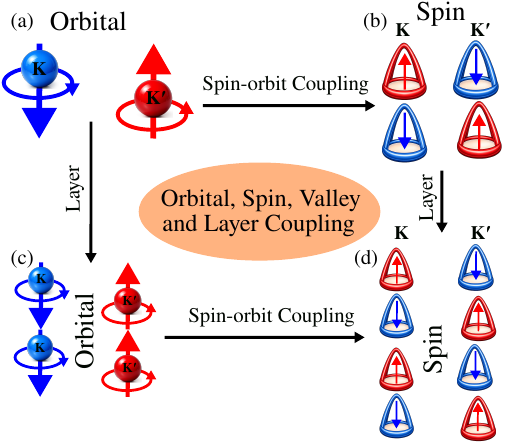}
\caption{A schematic illustration of orbital, spin, valley, and layer coupling.}
\label{fig:figure13}
\end{figure}

In summary, we demonstrate that Ta$_2$CS$_2$ provides an excellent platform for investigating the interplay among multiple degrees of freedom, viz., valley, orbital, spin, and layer, as schematically illustrated in Fig.~\ref{fig:figure13}. Using a combination of TB model based on Ta-$d$-only orbitals and complementary first-principles calculations, we establish the intricate interconnections among these degrees of freedom.

We show that inversion symmetry breaking in Ta$_2$CS$_2$ gives rise to non-zero orbital magnetic moments in momentum space. In particular, the states $|d_{x^2-y^2}\rangle \mp i|d_{xy}\rangle$, which are eigenstates of $L_z$ with eigenvalues $\pm 2\hbar$, lead to valley-contrasting orbital magnetic moments at the $K$ and $K'$ points of the BZ, as shown in Fig. \ref{fig:figure13}(a). The in-plane orbital texture around the $\Gamma$ point is Rashba texture which can be controlled by changing the direction of the polarization. Interestingly, both the in-plane and out-of-plane components of the orbital moment reverse upon switching the polarization direction, providing a knob to control orbital magnetic moments.


Upon inclusion of SOC, the orbital degrees of freedom couple to spin, leading to a spin texture as well. Consequently, the valley--orbital locking translates into spin--valley coupling, resulting in Zeeman-like spin splitting at the $K$ and $K'$ valleys with opposite spin polarization, as depicted in Fig. \ref{fig:figure13}(b). Furthermore, the ORE gives rise to a spin Rashba effect in the presence of SOC, generating an in-plane spin texture and momentum-dependent spin splitting around the $\Gamma$ point.

In the bilayer configuration, the layer degree of freedom introduces an additional level of complexity. We demonstrate the emergence of orbital-layer coupling, which, in the presence of SOC, further leads to spin-layer coupling, as shown schematically in Fig. \ref{fig:figure13}(c) and (d) respectively. This results in enhanced orbital moments at the valley points and additional layer-dependent spin splitting, highlighting the cooperative role of all four degrees of freedom.

Overall, our results establish the coexistence and mutual coupling of valley, orbital, spin, and layer degrees of freedom in Ta$_2$CS$_2$. This multi-degree-of-freedom interplay provides a versatile framework for tuning electronic structure and transport properties. In particular, phenomena such as the OHE and SHE are strongly influenced by this coupling, offering new avenues for controlling orbital and spin transport.


Looking ahead, the predicted orbital and spin textures can be probed experimentally using techniques such as circular dichroism measurements, which are sensitive to orbital character ~\cite{PhysRevLett.132.196402}; time-reversal dichroism in photoelectron angular distributions (TRDAD) for orbital texture~\cite{PhysRevLett.125.216404}; and spin-resolved ARPES ~\cite{doi:10.1126/science.1167733}, which enables the observation of spin texture. Similarly, the proposed OHE and SHE can be detected 
through orbital torque~\cite{PhysRevResearch.2.013177,Lee2021,Hayashi2023,PhysRevResearch.4.033037} and spin torque~\cite{RevModPhys.91.035004,Miron2011} measurements, respectively. In both cases, the transverse orbital and spin currents are injected into a ferromagnet, which exerts an orbital and spin torque, respectively. However, distinguishing the orbital torque arising from the OHE from the spin torque induced by the SHE remains a significant challenge~\cite{PhysRevResearch.2.033401}. Another approach to detecting the OHE involves measuring the orbital moments that accumulate at the surfaces of materials exhibiting the effect. This technique is analogous to the direct observation of the SHE in semiconductors~\cite{PhysRevLett.121.086602} and metals~\cite{PhysRevLett.119.087203}, as well as the valley Hall effect in two-dimensional materials~\cite{Mak2018}. It is interesting to point out here that the OHE-induced orbital moment has been successfully detected using the magneto-optical Kerr effect~\cite{Choi2023}.

Beyond Hall responses, the ORE identified here is expected to give rise to an orbital Edelstein effect \cite{1985JETP_61_133L,PhysRevResearch.3.013275,Salemi2019,PhysRevResearch.5.043294,Jo2024,PhysRevB.104.165403,PhysRevB.102.201403} under carrier doping, where an in-plane electric field induces a net orbital magnetization via a shift of the Fermi surface. In the presence of SOC, a corresponding spin Edelstein effect \cite{EDELSTEIN1990233,Han2018,PhysRevB.97.085417,PhysRevLett.114.166602} is also anticipated. Importantly, since both orbital and spin textures are coupled to the electric polarization, these effects are expected to be electrically switchable, providing a powerful mechanism for functional control.

By highlighting the strong coupling between multiple internal degrees of freedom, our work positions Ta$_2$CS$_2$, and more broadly two-dimensional non-centrosymmetric MXenes, as promising platforms for tunable quantum functionalities. We hope that these results will stimulate further theoretical and experimental efforts aimed at exploring multi-degree-of-freedom coupling and utilizing it for next-generation spin-orbitronic applications.

\begin{acknowledgments}
K.D. thanks the Council of Scientific and Industrial Research (CSIR) for support through a fellowship (File No. 09/080(1178)/2020-EMR-I). S.B. gratefully acknowledges financial support from the IRCC Seed Grant (Project Code: RD/0523-IRCCSH0-018), the INSPIRE Research Grant (Project Code: RD/0124-DST0030-002), and the ANRF PMECRG Grant (Project Code: RD/0125-ANRF000-019). I.D. thanks the Technical Research Center, Department of Science, and Technology for support. 
\end{acknowledgments}



\begin{thebibliography}{89}
\bibitem [{orb()}]{orbital_moment_qe}%
	\BibitemOpen
	\href@noop {} {}\bibinfo {note} {The orbital magnetic moment
		$\mathbf{M}^{\mathrm{orb}}(\mathbf{k})$ is calculated in \texttt{Quantum
			ESPRESSO} using the expression $\mathbf{M}^{\mathrm{orb}}(\mathbf{k}) =
		\sum_{n} \frac{e}{2\hbar} f_{n\mathbf{k}} \, \mathrm{Im} \left\langle
		\nabla_{\mathbf{k}} u_{n\mathbf{k}} \middle| \times \left( H_{\mathbf{k}} +
		\varepsilon_{n\mathbf{k}} - 2\varepsilon_F \right) \middle|
		\nabla_{\mathbf{k}} u_{n\mathbf{k}} \right\rangle ,$ where $f_{n\mathbf{k}}$
		is the Fermi--Dirac distribution function, $\varepsilon_{n\mathbf{k}}$ is the
		band energy, $H_{\mathbf{k}}$ is the $k$-dependent Hamiltonian, and
		$\varepsilon_F$ denotes the Fermi energy.}\BibitemShut {Stop}%
\end{thebibliography}

%
\clearpage
\onecolumngrid

\begin{center}
	{\Large \textbf{Supplemental Information}} \\
	\vspace{0.6 cm}
	{\large\textbf{Interplay of Valley, Orbital, Spin, and Layer Degrees of Freedom in Ta$_2$CS$_2$ MXene}}\\
	\vspace{0.3 cm}
	{{Kunal Dutta$^{1}$}, {Anupam Mondal$^{1}$}, {Sayantika Bhowal $^{2}$}, {Subhradip Ghosh$^{3}$}, and {Indra Dasgupta$^{1}$} }\\
	\vspace{0.4 cm}
	{$^{1}$ School of Physical Sciences, Indian Association for the Cultivation of Science, 2A and 2B Raja S.C. Mullick Road, Jadavpur, Kolkata 700032, India.\\
		$^{2}$ Department of Physics, Indian Institute of Technology Bombay, Mumbai 400076, India.\\
		$^{3}$ Department of Physics, Indian Institute of Technology Guwahati, Guwahati-781039, Assam, India.}

\end{center}

\setcounter{section}{0}
\setcounter{figure}{0}
\setcounter{table}{0}

\renewcommand{\thesection}{S\arabic{section}}
\renewcommand{\thefigure}{S\arabic{figure}}
\renewcommand{\thetable}{S\arabic{table}}

\section{Phonon Calculations for the Bilayer Structure}

\begin{figure}[ht]
	\includegraphics[height= 53 mm, width=78 mm,keepaspectratio]{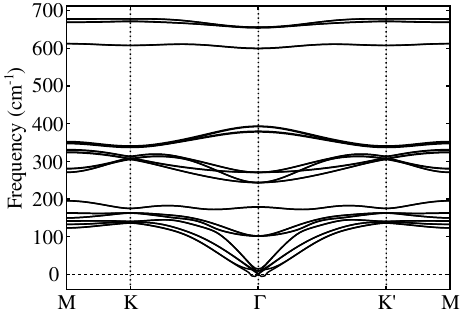}
	\caption{ Phonon band structure of bilayer Ta$_2$CS$_2$, plotted along the high-symmetry path $M\left(\frac{\pi}{a\sqrt{3}}, \frac{\pi}{a}, 0 \right)$–$K\left(0, \frac{4\pi}{3a}, 0 \right)$–$\Gamma\left(0, 0, 0\right)$–$K^{\prime}\left( 0, -\frac{4\pi}{3a}, 0 \right)$–$M\left( -\frac{\pi}{a\sqrt{3}}, -\frac{\pi}{a}, 0 \right)$.}
	\label{SFig:1}
\end{figure}

For any theoretically predicted material, it is essential to assess  its dynamical stability. In this work, two monolayers of Ta$_2$CS$_2$ are stacked on top of each other to form a bilayer configuration, referred to as AA stacking. Fig.~\ref{SFig:1} presents the phonon band structure of AA stacked bilayer Ta$_2$CS$_2$, calculated along the high-symmetry directions of the Brillouin zone. No imaginary phonon modes are observed in the phonon band structure, except for small U-shaped features in the first acoustic branch (ZA) near the zone center. These features do not indicate dynamical instability; instead, they are characteristic of layered two-dimensional systems~\cite{PhysRevX.11.041027,PhysRevB.95.165444,PhysRevLett.125.046402,PhysRevB.110.245421}.

\section{Electronic Band Structure from LMTO and NMTO Calculations}

\begin{figure}[ht]
	\includegraphics[height= 55 mm, width=81 mm,keepaspectratio]{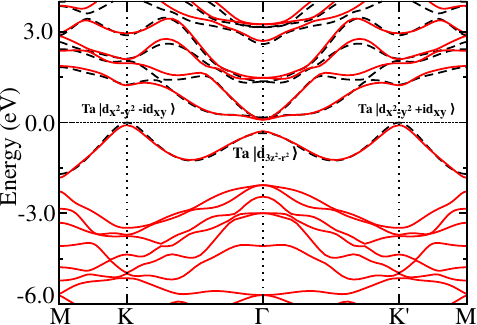}
	\caption{ LMTO band structure overlaid with the NMTO-downfolded tight-binding band structure, plotted along the high-symmetry path $M\left(\frac{\pi}{a\sqrt{3}}, \frac{\pi}{a}, 0 \right)$–$K\left(0, \frac{4\pi}{3a}, 0 \right)$–$\Gamma\left(0, 0, 0\right)$–$K^{\prime}\left( 0, -\frac{4\pi}{3a}, 0 \right)$–$M\left( -\frac{\pi}{a\sqrt{3}}, -\frac{\pi}{a}, 0 \right)$. The LMTO bands are shown by red solid lines, while the downfolded bands are represented by black dotted lines. }
	\label{SFig:2}
\end{figure}

To begin with, we consider the electronic band structure of Ta$_2$CS$_2$ in the 
absence of spin-orbit coupling (SOC), shown in Fig.~\ref{SFig:2} by the red solid lines, 
as obtained from linear muffin-tin orbital (LMTO) calculations. The compound 
exhibits insulating behavior. The origin of the semiconducting gap can be 
understood from simple electron-filling arguments.

The valence electron count for each constituent atom in Ta$_2$CS$_2$ is as 
follows: each Ta atom contributes five valence electrons, while C and S 
contribute four and six valence electrons, respectively. Consequently, 
Ta$_2$CS$_2$ has a total of 26 valence electrons per formula unit. These electrons 
fully occupy the 24 available states arising from the C-$s$, C-$p$, S-$s$, and 
S-$p$ orbitals, as illustrated in Fig.~\ref{SFig:2}. The remaining two electrons occupy a 
low-lying, isolated Ta $5d$ state associated with the prismatic coordination 
environment, thereby opening a band gap.

To construct a low-energy Hamiltonian we have downfolded the C-$s$, C-$p$, S-$s$, S-$p$ states retaining only Ta-d and Ta$_1$-d states in the basis. Using the $N$th-order muffin-tin 
orbital (NMTO) downfolding method.  The resulting downfolded band 
structure is shown by the black dotted lines in Fig.~\ref{SFig:2}, demonstrating 
excellent agreement with the full LMTO results in the low-energy window.


\section{Sign Changing Matrix Elements}

For both the downward- and upward-polarized structures of Ta$_2$CS$_2$, we perform a downfolding procedure by integrating out the C-$s$, C-$p$, S-$s$, S-$p$ states, retaining only the Ta-$d$ and Ta$_1$-$d$ orbitals in the basis. Using this reduced basis, we extract the hopping elements between the relevant orbitals and construct an effective low-energy tight-binding model. Among the extracted parameters, we identify a subset of hopping terms that change sign between the two polarization configurations. These sign-changing hopping parameters are summarized in Table~\ref{tab:Table1} and Table~\ref{tab:Table2} for the downward- and upward-polarized structures of Ta$_2$CS$_2$, respectively.

The hopping term 
$t^{\mathrm{Ta}\text{,}d\,\text{-}\,\mathrm{Ta}\text{,}d}_{[i,j,k]}$ denotes the 
hopping between Ta $d$ orbitals located in the same local (biprismatic) 
environment, including both nearest-neighbor and further-neighbor hopping 
processes. The integers $i$, $j$, and $k$ label the hopping processes between 
different unit cells along the corresponding lattice directions. The relative 
displacement vector associated with each hopping process is given by
$
\vec{r} = i\vec{a}_1 + j\vec{a}_2 + k\vec{a}_3 ,
$
where $\vec{a}_1$, $\vec{a}_2$, and $\vec{a}_3$ are the primitive lattice vectors. 
For convenience, the Fermi energy is set to $0.0$~eV for both polarization structures.

We observe that several hopping parameters between Ta $d$ orbitals in the 
bi-prismatic environment, including both nearest-neighbor and further-neighbor 
terms, undergo a sign change when transitioning between the two structural 
polarization structures. This sign reversal plays a crucial role in determining the orbital texture 
in these systems.

\begin{table*}[h]
	\centering
	\begin{tabular}{|c|c|c|c|c|}
		\hline\hline
		$i$ & $j$ & $k$& $\langle Ta\text{,}d|H| Ta\text{,}d\rangle$ & $t^{Ta\text{,}d-Ta\text{,}d}_{[i,j,k]}$ (meV) \\
		\hline
		$\quad$ 1 $\quad$&$\quad$ 1 $\quad$&$\quad$ 0 $\quad$& $\qquad \qquad \langle Ta\text{,}d_{x^2-y^2}|H| Ta{}\text{,}d_{z^2}\rangle \qquad \qquad$ & 590.49 \\
		1 & 0 & 0 & $\langle Ta\text{,}d_{x^2-y^2}|H| Ta{}\text{,}d_{z^2}\rangle$ & 590.49 \\
		-1 & 0 & 0 & $\langle Ta\text{,}d_{z^2}|H| Ta{}\text{,}d_{x^2-y^2}\rangle$ & 590.49 \\
		-1 & -1 & 0 & $\langle Ta\text{,}d_{z^2}|H| Ta{}\text{,}d_{x^2-y^2}\rangle$ & 590.49 \\
		1 & 0 & 0 & $\langle Ta\text{,}d_{x^2-y^2}|H| Ta{}\text{,}d_{xy}\rangle$ &  564.64 \\
		-1 & 0 & 0 & $\langle Ta\text{,}d_{xy}|H| Ta{}\text{,}d_{x^2-y^2}\rangle$ & 564.64 \\
		0 & 1 & 0 & $\langle Ta\text{,}d_{z^2}|H| Ta{}\text{,}d_{xy}\rangle$ &  432.66 \\
		0 & -1 & 0 & $\langle Ta\text{,}d_{xy}|H| Ta{}\text{,}d_{z^2}\rangle$ & 432.66 \\
		\hline\hline
	\end{tabular}
	\caption{Nearest-neighbor and further-neighbor hopping parameters, $t^{Ta\text{,}d-Ta\text{,}d}_{[i,j,k]}$ (in meV), for the downward-polarization structure of monolayer Ta$_2$CS$_2$.}
	\label{tab:Table1}
\end{table*}

\begin{table*}[h]
	\centering
	\begin{tabular}{|c|c|c|c|c|}
		\hline\hline
		$i$ & $j$ & $k$& $\langle Ta\text{,}d|H| Ta\text{,}d\rangle$ & $t^{Ta\text{,}d-Ta\text{,}d}_{[i,j,k]}$ (meV) \\
		\hline
		$\quad$ 1 $\quad$&$\quad$ 1 $\quad$&$\quad$ 0 $\quad$& $\qquad \qquad \langle Ta\text{,}d_{x^2-y^2}|H| Ta{}\text{,}d_{z^2}\rangle \qquad \qquad$ & -174.15 \\
		1 & 0 & 0 & $\langle Ta\text{,}d_{x^2-y^2}|H| Ta{}\text{,}d_{z^2}\rangle$ & -174.15 \\
		-1 & 0 & 0 & $\langle Ta\text{,}d_{z^2}|H| Ta{}\text{,}d_{x^2-y^2}\rangle$ & -174.15 \\
		-1 & -1 & 0 & $\langle Ta\text{,}d_{z^2}|H| Ta{}\text{,}d_{x^2-y^2}\rangle$ & -174.15 \\
		1 & 0 & 0 & $\langle Ta\text{,}d_{x^2-y^2}|H| Ta{}\text{,}d_{xy}\rangle$ &  -243.54 \\
		-1 & 0 & 0 & $\langle Ta\text{,}d_{xy}|H| Ta{}\text{,}d_{x^2-y^2}\rangle$ & -243.54 \\
		0 & 1 & 0 & $\langle Ta\text{,}d_{z^2}|H| Ta{}\text{,}d_{xy}\rangle$ &  -442.19 \\
		0 & -1 & 0 & $\langle Ta\text{,}d_{xy}|H| Ta{}\text{,}d_{z^2}\rangle$ & -442.19 \\
		\hline\hline
	\end{tabular}
	\caption{Nearest-neighbor and further-neighbor hopping parameters, $t^{Ta\text{,}d-Ta\text{,}d}_{[i,j,k]}$ (in meV), for the upward-polarization structure of monolayer Ta$_2$CS$_2$.}
	\label{tab:Table2}
\end{table*}

\end{document}